\documentclass[11pt,a4paper,english,superscriptaddress,aps,preprint]{revtex4}
\usepackage{xcolor}
\usepackage{amsfonts}
\usepackage{graphics}
\usepackage{graphicx}
\usepackage{hyperref}
\usepackage{slashed}
\usepackage{amsmath}
\usepackage{amssymb}
\makeatletter
\usepackage{babel}
\newcommand{\bea}{\begin{eqnarray}}
\newcommand{\eea}{\end{eqnarray}}

\begin{document}

\title{Spontaneous Symmetry Breaking in the Nonlocal Scalar QED}

\author{F. S. Gama}
\email{fisicofabricio@yahoo.com.br}
\affiliation{Departamento de F\'{\i}sica, Universidade Federal da Para\'{\i}ba\\
 Caixa Postal 5008, 58051-970, Jo\~ao Pessoa, Para\'{\i}ba, Brazil}

\author{J. R. Nascimento}
\email{jroberto@fisica.ufpb.br}
\affiliation{Departamento de F\'{\i}sica, Universidade Federal da Para\'{\i}ba\\
 Caixa Postal 5008, 58051-970, Jo\~ao Pessoa, Para\'{\i}ba, Brazil}

\author{A. Yu. Petrov}
\email{petrov@fisica.ufpb.br}
\affiliation{Departamento de F\'{\i}sica, Universidade Federal da Para\'{\i}ba\\
 Caixa Postal 5008, 58051-970, Jo\~ao Pessoa, Para\'{\i}ba, Brazil}

\author{P. J. Porf\'{i}rio}
\email{pporfirio89@gmail.com}
\affiliation{Departamento de F\'{\i}sica, Universidade Federal da Para\'{\i}ba\\
 Caixa Postal 5008, 58051-970, Jo\~ao Pessoa, Para\'{\i}ba, Brazil}

\begin{abstract}
We investigate the spontaneous symmetry breaking in a nonlocal version of the scalar QED. When the mass parameter $m^2$ satisfies the requirement $m^2>0$, we find that all fields, including the Nambu-Goldstone field, acquire a non-zero mass dependent on the nonlocal scales. On the other hand, when $m^2=0$, we find that the nonlocal corrections to the masses are very small and can be neglected.  
\end{abstract}

\maketitle

\section{Introduction}

The nonlocal theories actually acquire great attention within different contexts. The main reason for it consists in the natural ultraviolet finiteness of these theories which feeds the hopes that this methodology is appropriate for constructing the gravity model consistent at the quantum level while the known models for gravity are either non-renormalizable or involve the ghosts (negative energy states) whose presence breaks the unitarity. Another motivation for studying the nonlocal theories arises from phenomenology of elementary particles where already in 60s it was suggested that the methodology of nontrivial form factors describes correctly the elementary particles which naturally must be treated as finite-size objects. In a systematic manner, this methodology has been first presented in \cite{Efimov}. During a long time, the interest to nonlocal theories has been restricted by the phenomenological context, however, some suggestions within this context are very interesting, for example, it deserves to mention that the nonlocality finds applications within studies of confinement, see f.e. \cite{Efimov96}.

A strong growing of interest to nonlocal theories in general quantum field theory context began after formulation of the concept of the space-time noncommutativity \cite{SW}. It gave origin to formulating of noncommutative theories on the base of the Moyal product which is known to be one of the simplest way to introduce the nonlocality in general field theory context. However, one of the key problems within this methodology, which has not been solved up to now, is the development of the Moyal-like formulation for gravity. Therefore, other manners to implement the nonlocality, with the most popular among them is based on introducing of nonpolynomial functions $f(\Box)$ to the classical action, became more important. The seminal role in this direction has been played by the paper \cite{BMS} where it was argued, with use of stringy motivation which naturally involve higher derivatives, that infinite-derivative gravity theories do not involve ghosts and allow to eliminate the initial singularity (see also \cite{KMS}). Therefore, the role of the nonlocality within the gravity context seems to be the fundamental one where the nonlocal extension allows to achieve the super-renormalizability \cite{Modesto}. At the same time, the problem of nonlocal extensions for other field theory models also becomes interesting. Indeed, from one side, these models become a convenient laboratory for studying of quantum impacts of nonlocality whether the nonlocal gravity apparently will be an extremely complicated theory at the quantum level. From other side, the elimination of ultraviolet divergences due to the nonlocality naturally improves the qualities of these theories (some interesting results for nonlocal non-gravitational theories can be found in \cite{Modesto1,Bri}). At the same time, it is important to mention that many known statements of the quantum field theories are well proved namely for the local theories. The typical example of such a statement is the Goldstone theorem. It was explicitly demonstrated in \cite{ourSSB} that in the Moyal-like theories it is satisfied only in certain cases. Hence the natural problem consists in study of the consistency of this theorem in the nonlocal theories based on nonpolynomial $f(\Box)$ functions. This is the problem we consider here.

The structure of this paper looks like follows. In the section 2, we formulate the nonlocal scalar QED on the classical level. In the section 3, we study the symmetry breaking in this theory on the tree level, and in the section 4 -- on the one-loop level. The section 5 is the summary where we discuss our results.


\section{Nonlocal Scalar QED}

Let us start our study with a nonlocal version of the scalar quantum electrodynamics. We define the classical Lagrangian for this theory as: 
\bea
\label{NLSQED}
{\mathcal L}=-\frac{1}{4}F^{\mu\nu}e^{\frac{\Box}{\Lambda^2_A}}F_{\mu\nu}-\frac{1}{2}\bigg[\phi^*e^{\frac{\Box_c}{\Lambda^2_\phi}}\left(\Box_c-m^2\right)\phi+h.c.\bigg]-\frac{\lambda}{3!}\left|\phi\right|^4 \ ,
\eea
where $\Box_c=D^\mu D_\mu$ is the covariant d'Alembertian operator, $D_\mu\phi=\partial_\mu\phi+iqA_\mu\phi$, and $F_{\mu\nu}$ is the usual electromagnetic field tensor. Additionally, the parameters $\Lambda_A$ and $\Lambda_\phi$ are mass scales in which the nonlocal contributions are significant. Note that the local scalar QED can be formally recovered by taking the limits $\Lambda_A\rightarrow\infty$ and $\Lambda_\phi\rightarrow\infty$ in Eq. (\ref{NLSQED}). Finally, we assume that the nonlocal operators are given by the following infinite series \cite{Efimov}:
\bea
\label{series}
e^{\frac{\Box}{\Lambda^2_A}}=\sum_{n=0}^{\infty}\frac{1}{n!}\frac{\Box^n}{\Lambda_A^{2n}} \ ; \ e^{\frac{\Box_c}{\Lambda^2_\phi}}=\sum_{n=0}^{\infty}\frac{1}{n!}\frac{\Box_c^n}{\Lambda_\phi^{2n}} \ .
\eea
Put in another way, the $e^{\frac{\Box}{\Lambda^2_A}}$ and $e^{\frac{\Box_c}{\Lambda^2_\phi}}$ terms can be thought as shorthand notations for the infinite series above.

In order to gain a further understanding of the model (\ref{NLSQED}), let us calculate the propagators of the gauge and scalar fields. Since the Lagrangian (\ref{NLSQED}) is invariant under local $U(1)$ transformations, it is necessary to fix the gauge. Thus, we add to (\ref{NLSQED}) the following gauge-fixing term \cite{BO}
\bea
\label{gaugefixing}
\mathcal{L}_{GF}=-\frac{1}{2\xi}\Big(e^{\frac{\Box}{2\Lambda^2_A}}\partial_\mu A^\mu\Big)^2 \ ,
\eea
which is a natural nonlocal generalization of the standard Fermi gauges, so that (\ref{gaugefixing}) does not explicitly break the global $U(1)$ symmetry and the ghosts associated with this gauge fixing decouple.

It follows from (\ref{NLSQED}) and (\ref{gaugefixing}) that the propagators of the gauge and scalar fields are given by
\bea
\label{prop}
\left\langle A_\mu(-p)A_\nu(p)\right\rangle=-\frac{i \ e^{\frac{p^2}{\Lambda^2_A}}}{p^2}\Big[\left(P_T\right)_{\mu\nu}+\xi\left(P_L\right)_{\mu\nu}\Big] \ ; \ \left\langle \phi^*(-p)\phi(p)\right\rangle=\frac{i \ e^{\frac{p^2}{\Lambda^2_\phi}}}{p^2+m^2} \ ,
\eea
where we wrote the gauge propagator in terms of the projection operators
\bea
\label{proj}
\left(P_T\right)_{\mu\nu}=\eta_{\mu\nu}-\frac{\partial_\mu\partial_\nu}{\Box} \ ; \ \left(P_L\right)_{\mu\nu}=\frac{\partial_\mu\partial_\nu}{\Box} \ .
\eea
We note from (\ref{prop}) that the introduction of the nonlocal terms (\ref{series}) into the scalar QED improves the ultraviolet behavior of the theory without introducing unwanted degrees of freedom (ghosts) \cite{BMS}. Indeed, after a Wick rotation to the Euclidean space $p^0\rightarrow ip_E^0$, we obtain $p^2\rightarrow-p_E^2$, this implies that the ultraviolet modes will be suppressed by the Gaussian functions carried by the propagators. At the same time, since the exponential of an entire function is an entire function with no zeros on the whole complex plane, this ensure that the theory (\ref{NLSQED}) does not contain extra degrees of freedom as compared to the local scalar QED.

\section{Tree-level Breaking of Symmetry}
\label{TLBS}

Our goal in this section is to study the process of spontaneous symmetry breaking in the nonlocal scalar QED at the tree level. In order to achieve this, we have to make the assumption that $m^2>0$ \cite{Andreassen}. In the case where $m^2>0$, the theory given by the Lagrangian (\ref{NLSQED}) can be called the nonlocal Abelian Higgs model (NLAHM). For this model, we will calculate the dispersion relations associated with the free-field equations and obtain the masses of the fields.

From (\ref{NLSQED}), we can infer that the tree-level potential is given by
\bea
\label{treepot}
V(\phi)=-m^2\left|\phi\right|^2+\frac{\lambda}{3!}\left|\phi\right|^4 \ .
\eea
For $m^2>0$, the minimum of $V(\phi)$ occurs when $\left|\phi\right|^2=\frac{3m^2}{\lambda}\equiv\frac{v^2}{2}$, so that the $U(1)$ symmetry is spontaneously broken. Instead of dealing with the complex field $\phi$, it is convenient to write $\phi$ in terms of real fields $\sigma$ and $\pi$ which have zero vacuum expectation values. Thus, we choose \cite{AH}
\bea
\label{exp}
\phi(x)=\left(\frac{v+\sigma(x)}{\sqrt{2}}\right)e^{i\frac{\pi(x)}{v}} \ .
\eea
Substituting (\ref{exp}) into (\ref{NLSQED}), we obtain
\bea
\label{reallag}
{\mathcal L}&=&-\frac{1}{4}F^{\mu\nu}e^{\frac{\Box}{\Lambda^2_A}}F_{\mu\nu}-\frac{1}{4}\bigg[v^2e^{-i\frac{\pi}{v}}e^{\frac{\Box_c}{\Lambda^2_\phi}}\left(\Box_c-m^2\right)e^{i\frac{\pi}{v}}+ve^{-i\frac{\pi}{v}}e^{\frac{\Box_c}{\Lambda^2_\phi}}\left(\Box_c-m^2\right)\big(\sigma e^{i\frac{\pi}{v}}\big)\nonumber\\
&+&v\sigma e^{-i\frac{\pi}{v}}e^{\frac{\Box_c}{\Lambda^2_\phi}}\left(\Box_c-m^2\right)e^{i\frac{\pi}{v}}+\sigma e^{-i\frac{\pi}{v}}e^{\frac{\Box_c}{\Lambda^2_\phi}}\left(\Box_c-m^2\right)\big(\sigma e^{i\frac{\pi}{v}}\big)+h.c.\bigg]
-\frac{\lambda}{4!}\left(v+\sigma\right)^4 \ .
\eea
Since we want to obtain the free-field equations and the nonlocal operator in the scalar sector involves a covariant d'Alembertian $\Box_c$, we have to use Eq. (\ref{series}) and calculate each term of the series up to the second order in the fields. For example,
\bea
v^2e^{-i\frac{\pi}{v}}e^{\frac{\Box_c}{\Lambda^2_\phi}}\Box_ce^{i\frac{\pi}{v}}&=&v^2e^{-i\frac{\pi}{v}}\bigg(\Box_c+\frac{1}{\Lambda^2_\phi}\Box_c^2+\frac{1}{2\Lambda^4_\phi}\Box_c^3+\cdots\bigg)e^{i\frac{\pi}{v}}\nonumber\\
&\approx&v^2\bigg\{\frac{1}{v^2}\pi\Box\pi-\frac{2q}{v}A^\mu\partial_\mu\pi-q^2A^\mu A_\mu\nonumber\\
&+&\frac{1}{\Lambda^2_\phi}\left[\frac{1}{v^2}\pi\Box^2\pi-\frac{2q}{v}A^\mu\Box\partial_\mu\pi-q^2A^\mu\partial_\mu\partial_\nu A^\nu\right]\nonumber\\
&+&\frac{1}{2\Lambda^4_\phi}\left[\frac{1}{v^2}\pi\Box^3\pi-\frac{2q}{v}A^\mu\Box^2\partial_\mu\pi-q^2A^\mu\Box\partial_\mu\partial_\nu A^\nu\right]+\cdots\bigg\} \ .
\eea
Thus, we can show with the aid of (\ref{proj}) that
\bea
\label{sexp1}
v^2e^{-i\frac{\pi}{v}}e^{\frac{\Box_c}{\Lambda^2_\phi}}\Box_ce^{i\frac{\pi}{v}}\approx\pi e^{\frac{\Box}{\Lambda^2_\phi}}\Box\pi-2qvA^\mu e^{\frac{\Box}{\Lambda^2_\phi}}\partial_\mu\pi-q^2v^2A^\mu\Big[\left(P_T\right)_{\mu\nu}+e^{\frac{\Box}{\Lambda^2_\phi}}\left(P_L\right)_{\mu\nu}\Big]A^\nu \ .
\eea
Similarly, we can also show that
\bea
v^2e^{-i\frac{\pi}{v}}e^{\frac{\Box_c}{\Lambda^2_\phi}}e^{i\frac{\pi}{v}}&\approx& v^2+\pi\Big(e^{\frac{\Box}{\Lambda^2_\phi}}-1\Big)\pi-2qvA^\mu \frac{\Big(e^{\frac{\Box}{\Lambda^2_\phi}}-1\Big)}{\Box}\partial_\mu\pi-q^2v^2A^\mu\Big[\frac{1}{\Lambda^2_\phi}\left(P_T\right)_{\mu\nu}\nonumber\\
&+&\frac{\Big(e^{\frac{\Box}{\Lambda^2_\phi}}-1\Big)}{\Box}\left(P_L\right)_{\mu\nu}\Big]A^\nu \ ;\\
ve^{-i\frac{\pi}{v}}e^{\frac{\Box_c}{\Lambda^2_\phi}}\Box_c\big(\sigma e^{i\frac{\pi}{v}}\big)&\approx&-i\pi e^{\frac{\Box}{\Lambda^2_\phi}}\Box\sigma+iqvA^\mu e^{\frac{\Box}{\Lambda^2_\phi}}\partial_\mu\sigma \ ;\\
ve^{-i\frac{\pi}{v}}e^{\frac{\Box_c}{\Lambda^2_\phi}}\big(\sigma e^{i\frac{\pi}{v}}\big)&\approx&v\sigma-i\pi \Big(e^{\frac{\Box}{\Lambda^2_\phi}}-1\Big)\sigma+iqvA^\mu \frac{\Big(e^{\frac{\Box}{\Lambda^2_\phi}}-1\Big)}{\Box}\partial_\mu\sigma \ ;\\
v\sigma e^{-i\frac{\pi}{v}}e^{\frac{\Box_c}{\Lambda^2_\phi}}\Box_c e^{i\frac{\pi}{v}}&\approx&i\sigma e^{\frac{\Box}{\Lambda^2_\phi}}\Box\pi+iqv\sigma e^{\frac{\Box}{\Lambda^2_\phi}}\partial_\mu A^\mu \ ;\\
\label{sexp2}
v\sigma e^{-i\frac{\pi}{v}}e^{\frac{\Box_c}{\Lambda^2_\phi}}e^{i\frac{\pi}{v}}&\approx&v\sigma+i\sigma \Big(e^{\frac{\Box}{\Lambda^2_\phi}}-1\Big)\pi+iqv\sigma \frac{\Big(e^{\frac{\Box}{\Lambda^2_\phi}}-1\Big)}{\Box}\partial_\mu A^\mu \ .
\eea
Therefore, by substituting (\ref{sexp1}-\ref{sexp2}) into (\ref{reallag}) and using the definition of $v$ to simplify some of the terms, we get
\bea
\label{ffl}
{\mathcal L}&=&\frac{1}{2}A^\mu\left\{\bigg[e^{\frac{\Box}{\Lambda^2_A}}\Box+q^2v^2\bigg(1-\frac{m^2}{\Lambda^2_\phi}\bigg)\bigg]\left(P_T\right)_{\mu\nu}+q^2v^2\bigg[e^{\frac{\Box}{\Lambda^2_\phi}}-\frac{m^2}{\Box}\Big(e^{\frac{\Box}{\Lambda^2_\phi}}-1\Big)\bigg]\left(P_L\right)_{\mu\nu}\right\}A^\nu\nonumber\\
&-&\frac{1}{2}\sigma\Big[e^{\frac{\Box}{\Lambda^2_\phi}}(\Box-m^2)+3m^2\Big]\sigma-\frac{1}{2}\pi\Big[e^{\frac{\Box}{\Lambda^2_\phi}}(\Box-m^2)+m^2\Big]\pi+qvA^\mu\bigg[e^{\frac{\Box}{\Lambda^2_\phi}}-\frac{m^2}{\Box}\nonumber\\
&\times&\Big(e^{\frac{\Box}{\Lambda^2_\phi}}-1\Big)\bigg]\partial_\mu\pi+\frac{3m^4}{2\lambda}+{\mathcal L}_{int} \ ,
\eea
where ${\mathcal L}_{int}$ is the interaction Lagrangian.

While the vacuum breaks the symmetry, the above Lagrangian remains invariant under the following local gauge transformations \cite{Das1}:
\bea
\delta\sigma(x)=0 \ ; \ \delta\pi(x)=-v\alpha(x) \ ; \ \delta A_\mu(x)=\frac{1}{q}\partial_\mu\alpha(x) \ .
\eea
In order to understand the physical content of the free-field lagrangian (\ref{ffl}), it is convenient to use the gauge freedom to eliminate the mixing term between $\pi$ and $A_\mu$. For the purposes of this paper, we choose the Lorentz gauge $\partial_\mu A^\mu=0$, so that the field equations are given by
\bea
\bigg[e^{\frac{\Box}{\Lambda^2_A}}\Box+q^2v^2\bigg(1-\frac{m^2}{\Lambda^2_\phi}\bigg)\bigg]A^\mu=0 \ ; \Big[e^{\frac{\Box}{\Lambda^2_\phi}}(\Box-m^2)+3m^2\Big]\sigma=0 \ ; \Big[e^{\frac{\Box}{\Lambda^2_\phi}}(\Box-m^2)+m^2\Big]\pi=0 .
\eea
These field equations do admit plane-wave solutions $e^{\pm ipx}$ with the following dispersion relations
\bea
\label{dr}
e^{-\frac{p^2}{\Lambda^2_A}}p^2=q^2v^2\bigg(1-\frac{m^2}{\Lambda^2_\phi}\bigg) \ ; \ e^{-\frac{p^2}{\Lambda^2_\phi}}(p^2+m^2)=3m^2 \ ; \ e^{-\frac{p^2}{\Lambda^2_\phi}}(p^2+m^2)=m^2 \ .
\eea
Note that the dispersion relations are transcendental equations. Fortunately, Eqs. (\ref{dr}) can be analytically solved with the help of the Lambert-$W$ function. This function $W(z)$ is defined to be the multivalued inverse of the function $f(z)=ze^z$ \cite{CGHJK}. Therefore, the general solutions of the dispersion relations (\ref{dr}) are given by
\bea
\label{sol}
p^2=m^2_i \ ; \ i=A,\sigma,\pi \ ,
\eea
where
\bea
\label{complexmasses}
&&m^2_A=-\Lambda_A^2 W_k\left[-\frac{q^2v^2}{\Lambda_A^2}\left(1-\frac{m^2}{\Lambda_\phi^2}\right)\right] \ ; \ m^2_\sigma=-m^2-\Lambda_\phi^2 W_k\left[-\frac{3m^2}{\Lambda_\phi^2}e^{-\frac{m^2}{\Lambda_\phi^2}}\right] \ ;\nonumber \\
&&m^2_\pi=-m^2-\Lambda_\phi^2 W_k\left[-\frac{m^2}{\Lambda_\phi^2}e^{-\frac{m^2}{\Lambda_\phi^2}}\right] \ ,
\eea
for all $k\in\mathbb{Z}$. Here, the index $k$ denotes the branches of $W(z)$ \cite{VJC}. It is worth to point out that most of these branches are unphysical due to the fact that $W(z)$ is a multivalued complex function. However, if $z\in\mathbb{R}$ and $z\geq-e^{-1}$, then there are two possible real branches of $W(z)$: the upper branch satisfying $W(z)\geq-1$, which is labelled by $k=0$, and the bottom branch satisfying $W(z)\leq-1$, which is labelled by $k=-1$ (see Fig. 1). Therefore, it follows from this discussion and Eq. (\ref{complexmasses}) that $m^2_i\in\mathbb{R}$ only if
\bea
\label{constraint1}
\frac{q^2v^2}{\Lambda_A^2}\left(1-\frac{m^2}{\Lambda_\phi^2}\right)\leq\frac{1}{e} \ ; \ \frac{3m^2}{\Lambda_\phi^2}e^{-\frac{m^2}{\Lambda_\phi^2}}\leq\frac{1}{e} \ .
\eea

\begin{figure}[!h]
\begin{center}
\includegraphics[angle=0,scale=0.4]{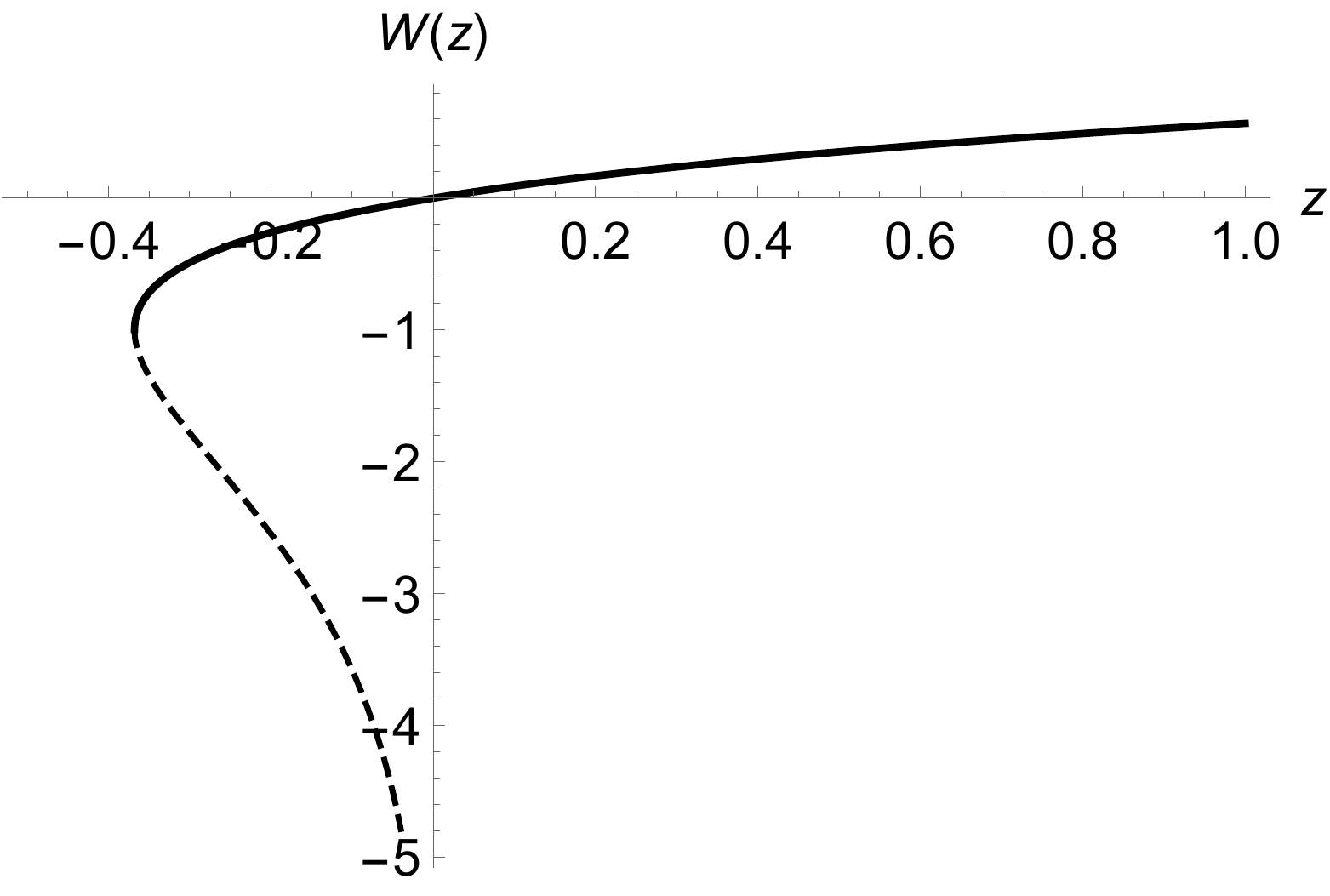}
\end{center}
\caption{The two real branches of $W(z)$. The solid line represents the upper branch $W_0(z)$, while the dashed line represents the bottom branch $W_{-1}(z)$.}
\end{figure}

We notice in Fig. 1 that $W(z)$ is a double-valued real function on $-e^{-1}\leq z<0$ and a single-valued real function on $z\geq0$ \cite{CM}. Moreover, we also notice that the bottom branch $W_{-1}(z)$ has a negative singularity for $z\rightarrow 0^{-}$, while the upper branch $W_{0}(z)$ is real-analytic at $z=0$. This observation allows us to conclude that only $W_{0}(z)$ is physically acceptable, due to the fact that all dependence on $\Lambda_A$ and $\Lambda_\phi$ must decouple from the masses as $\Lambda_A\rightarrow\infty$ and $\Lambda_\phi\rightarrow\infty$ in (\ref{complexmasses}).

Additionally, since the masses $m^2_i$ must be non-negative numbers, then the following constraints must also be satisfied:
\bea
\label{constraint2}
W_0\left[-\frac{q^2v^2}{\Lambda_A^2}\left(1-\frac{m^2}{\Lambda_\phi^2}\right)\right]\leq 0 \ ; \ W_0\left[-\frac{3m^2}{\Lambda_\phi^2}e^{-\frac{m^2}{\Lambda_\phi^2}}\right]\leq-\frac{m^2}{\Lambda_\phi^2} \ ; \ W_0\left[-\frac{m^2}{\Lambda_\phi^2}e^{-\frac{m^2}{\Lambda_\phi^2}}\right]\leq-\frac{m^2}{\Lambda_\phi^2} \ .
\eea
All constraints (\ref{constraint1}) and (\ref{constraint2}) are satisfied in the particular case where the nonlocal scales $\Lambda_A$ and $\Lambda_\phi$ are much larger than all the other mass parameters $m$ and $v$.

Therefore, we can state that, given the constraints in (\ref{constraint1}) and (\ref{constraint2}), in a process of spontaneous symmetry breaking, the fields acquire the following non-zero masses:
\bea
\label{ma}
m^2_A&=&-\Lambda_A^2 W_0\left[-\frac{q^2v^2}{\Lambda_A^2}\left(1-\frac{m^2}{\Lambda_\phi^2}\right)\right]\approx q^2v^2\left(1-\frac{m^2}{\Lambda_\phi^2}+\frac{q^2v^2}{\Lambda_A^2}-\frac{2m^2q^2v^2}{\Lambda_\phi^2\Lambda_A^2}+\frac{3q^4v^4}{2\Lambda_A^4}-\cdots\right) ;\\
\label{ms}
m^2_\sigma&=&-m^2-\Lambda_\phi^2 W_0\left[-\frac{3m^2}{\Lambda_\phi^2}e^{-\frac{m^2}{\Lambda_\phi^2}}\right]\approx 2m^2\left(1+\frac{3m^2}{\Lambda_\phi^2}+\frac{12m^4}{\Lambda_\phi^4}+\cdots\right) ;\\
\label{mp}
m^2_\pi&=&-m^2-\Lambda_\phi^2 W_0\left[-\frac{m^2}{\Lambda_\phi^2}e^{-\frac{m^2}{\Lambda_\phi^2}}\right]\xrightarrow{\Lambda_\phi\rightarrow\infty}0 \ ,
\eea
where all masses are dependent on the nonlocal scales, but $m^2_\pi$ is not analytic at $\Lambda_\phi\rightarrow\infty$.

Finally, on physical grounds, we can conclude that there is a unique solution (\ref{sol}) for each dispersion relation (\ref{dr}), where the masses of the fields are given by (\ref{ma}-\ref{mp}). This implies that the NLAHM contains the same number of degrees of freedom as the original local model. On the other hand, it is important to note that the Nambu-Goldstone field also acquired a non-zero mass dependent on the nonlocal scale. Therefore, we find that the Goldstone theorem is not valid for the NLAHM at the tree level. It is worth pointing out that this conclusion is based on the assumption that $m^2>0$. Thus, it would be interesting to check if the Goldstone theorem is valid for (\ref{NLSQED}) in the case where $m^2=0$.

\section{Quantum Breaking of Symmetry}
\label{QBS}

If we set $m^2=0$ in Eq. (\ref{treepot}), there will be no spontaneous symmetry breaking at the tree level. However, the symmetry can still be spontaneously broken as a result of quantum corrections to the classical potential (\ref{treepot}). This is, of course, the well-known Coleman-Weinberg mechanism \cite{CW,Weinberg}.  Thus, our goal in this section is to calculate the effective potential to one-loop order for the nonlocal massless scalar QED and use it to find the modified dispersion relations for the fields.

In order to calculate the one-loop correction to the effective potential, we will employ the background-field method \cite{BOS}. Following this approach, we make the shifts $A_\mu\rightarrow\hat{A}_\mu+A_\mu$ and $\phi\rightarrow\hat{\phi}+\phi$, where $\hat{A_\mu}$ and $\hat{\phi}$ are background fields, while $A_\mu$ and $\phi$ are quantum ones. We assume that the background fields are subject to the constraints $\hat{A}_\mu=0$ and $\partial_\mu\hat{\phi}=0$. Thus, it follows from (\ref{NLSQED}) and (\ref{gaugefixing}) that
\bea
{\mathcal L}+\mathcal{L}_{GF}&=&-\frac{1}{4}F^{\mu\nu}e^{\frac{\Box}{\Lambda^2_A}}F_{\mu\nu}-\frac{1}{2\xi}\Big(e^{\frac{\Box}{2\Lambda^2_A}}\partial_\mu A^\mu\Big)^2-\frac{1}{2}\bigg[\hat{\phi}^*e^{\frac{\Box_c}{\Lambda^2_\phi}}\left(\Box_c-m^2\right)\hat{\phi}+\hat{\phi}^*e^{\frac{\Box_c}{\Lambda^2_\phi}}\big(\Box_c-m^2\big)\phi \ \ \ \  \nonumber\\
&+&\phi^*e^{\frac{\Box_c}{\Lambda^2_\phi}}\left(\Box_c-m^2\right)\hat{\phi}+\phi^*e^{\frac{\Box_c}{\Lambda^2_\phi}}\left(\Box_c-m^2\right)\phi+h.c.\bigg]-\frac{\lambda}{3!}\big|\hat{\phi}+\phi\big|^4 \ .
\eea
For the sake of generality, we are considering $m^2\neq0$, although we will take $m^2=0$ later. 

In the one-loop approximation, we have to keep only the quadratic terms in the quantum fields. Thus, this approximation allows us to show that
\bea
\hat{\phi}^*e^{\frac{\Box_c}{\Lambda^2_\phi}}\Box_c\hat{\phi}&\approx&-q^2\big|\hat{\phi}\big|^2A^\mu\Big[\left(P_T\right)_{\mu\nu}+e^{\frac{\Box}{\Lambda^2_\phi}}\left(P_L\right)_{\mu\nu}\Big]A^\nu \ ;\\
\hat{\phi}^*e^{\frac{\Box_c}{\Lambda^2_\phi}}\hat{\phi}&\approx& -q^2\big|\hat{\phi}\big|^2A^\mu\Big[\frac{1}{\Lambda^2_\phi}\left(P_T\right)_{\mu\nu}+\frac{\Big(e^{\frac{\Box}{\Lambda^2_\phi}}-1\Big)}{\Box}\left(P_L\right)_{\mu\nu}\Big]A^\nu \ ;\\
\hat{\phi}^*e^{\frac{\Box_c}{\Lambda^2_\phi}}\Box_c\phi&\approx&iq\hat{\phi}^*A^\mu e^{\frac{\Box}{\Lambda^2_\phi}}\partial_\mu\phi \ ; \ \hat{\phi}^*e^{\frac{\Box_c}{\Lambda^2_\phi}}\phi \ \approx \  iq\hat{\phi}^*A^\mu \frac{\Big(e^{\frac{\Box}{\Lambda^2_\phi}}-1\Big)}{\Box}\partial_\mu\phi \ ;\\
\phi^*e^{\frac{\Box_c}{\Lambda^2_\phi}}\Box_c\hat{\phi}&\approx&iq\hat{\phi}\phi^* e^{\frac{\Box}{\Lambda^2_\phi}}\partial_\mu A^\mu \ ; \
\phi^*e^{\frac{\Box_c}{\Lambda^2_\phi}}\hat{\phi} \ \approx \ iq\hat{\phi}\phi^* \frac{\Big(e^{\frac{\Box}{\Lambda^2_\phi}}-1\Big)}{\Box}\partial_\mu A^\mu \ ,
\eea
As a result, the quadratic lagrangian of quantum fields looks like
\bea
\label{quadlag}
\mathcal{L}_2&=&\frac{1}{2}A^\mu\left\{\bigg[e^{\frac{\Box}{\Lambda^2_A}}\Box+2q^2\big|\hat{\phi}\big|^2\bigg(1-\frac{m^2}{\Lambda^2_\phi}\bigg)\bigg]\left(P_T\right)_{\mu\nu}+\bigg[\xi^{-1}e^{\frac{\Box}{\Lambda^2_A}}\Box+2q^2\big|\hat{\phi}\big|^2f(\Box)\bigg]\left(P_L\right)_{\mu\nu}\right\}A^\nu \nonumber\\
&-&\phi^*e^{\frac{\Box_c}{\Lambda^2_\phi}}\left(\Box_c-m^2\right)\phi+iqA^\mu\Big[\hat{\phi}f(\Box)\partial_\mu\phi^*-\hat{\phi}^*f(\Box)\partial_\mu\phi\Big]-\frac{\lambda}{3!}\big(\hat{\phi}^2\phi^{*2}+2\big|\hat{\phi}\big|^2\big|\phi\big|^2+h.c.\big)\ , \ \ \ \ \
\eea
where $f(\Box)\equiv e^{\frac{\Box}{\Lambda^2_\phi}}-\frac{m^2}{\Box}(e^{\frac{\Box}{\Lambda^2_\phi}}-1)$.

Instead of dealing with one complex field, it is easier to deal with two real fields. Thus, let us define $\phi(x)=2^{-1/2}[\phi_1(x)+i\phi_2(x)]$ and rewrite (\ref{quadlag}) as 
\bea
\label{L2}
\mathcal{L}_2=\frac{1}{2}
\left(\begin{array}{cc}
A^\mu & \phi_a
\end{array}\right)
\left(\begin{array}{cc}
\widehat{A} & \widehat{B}\ \\
\widehat{C} & \widehat{D} 
\end{array}\right)
\left(\begin{array}{c}
A^\nu \\
\phi_b
\end{array}\right) \ ,
\eea
where
\bea
\widehat{A}&=&\bigg[e^{\frac{\Box}{\Lambda^2_A}}\Box+q^2\hat{\phi}^2\bigg(1-\frac{m^2}{\Lambda^2_\phi}\bigg)\bigg]\left(P_T\right)_{\mu\nu}+\bigg[\xi^{-1}e^{\frac{\Box}{\Lambda^2_A}}\Box+q^2\hat{\phi}^2f(\Box)\bigg]\left(P_L\right)_{\mu\nu} \ ;\\
\widehat{B}&=&q\hat{\phi}_a\varepsilon_{ab}f(\Box)\partial_\mu \ ; \ \widehat{C} \ = \ q\varepsilon_{ab}\hat{\phi}_bf(\Box)\partial_\nu \ ;\\
\widehat{D}&=&-\Big[e^{\frac{\Box}{\Lambda^2_\phi}}\left(\Box-m^2\right)+\frac{\lambda}{6}\hat{\phi}^2\Big]\delta_{ab}-\frac{\lambda}{3}\hat{\phi}_a\hat{\phi}_b \ .
\eea
Moreover, $\hat{\phi}^2\equiv\hat{\phi}^2_1+\hat{\phi}^2_2$, the index $a$ runs from $1$ to $2$, and $\varepsilon_{ab}$ is the Levi-Civita symbol.

By integrating out the quantum fields in Eq. (\ref{L2}), it is possible to show that the one-loop contribution to the effective potential is given by \cite{Das2}
\bea
\label{1loopEP}
V^{(1)}(\hat{\phi})=-\frac{i}{2\Omega}\textrm{Tr}\ln\left(\begin{array}{cc}
\widehat{A} & \widehat{B}\ \\
\widehat{C} & \widehat{D} 
\end{array}\right) \ , 
\eea
where $\textrm{Tr}$ represents the trace over the matrix, Lorentz, and spacetime indices. The factor $\Omega$ denotes the spacetime volume.

We can split the above trace into two pieces by using the following matrix identity:
\bea
\label{ident}
\textrm{Tr}\ln\left(\begin{array}{cc}
\widehat{A}&\widehat{B}\\
\widehat{C} & \widehat{D}\\
\end{array}\right)=\textrm{Tr}\ln\widehat{D}+\textrm{Tr}\ln\left(\widehat{A}-\widehat{B}\widehat{D}^{-1}\widehat{C}\right) \ .
\eea
In order to use this identity, we first need to determine the inverse of $\widehat{D}$. Thus, we get
\bea
\label{inverse}
D^{-1}=-\Big[e^{\frac{\Box}{\Lambda^2_\phi}}(\Box-m^2)+\frac{\lambda}{6}\hat{\phi}^2\Big]^{-1}\bigg\{\delta_{ab}-\frac{\lambda}{3}\Big[e^{\frac{\Box}{\Lambda^2_\phi}}(\Box-m^2)+\frac{\lambda}{2}\hat{\phi}^2\Big]^{-1}\hat{\phi}_a\hat{\phi}_b\bigg\} \ .
\eea
Therefore, it follows from (\ref{1loopEP}-\ref{inverse}) that
\bea
\label{EP}
V^{(1)}(\hat{\phi})&=&-\frac{i}{2\Omega}\textrm{Tr}\ln\bigg\{-\Big[e^{\frac{\Box}{\Lambda^2_\phi}}\left(\Box-m^2\right)+\frac{\lambda}{6}\hat{\phi}^2\Big]\delta_{ab}-\frac{\lambda}{3}\hat{\phi}_a\hat{\phi}_b\bigg\}-\frac{i}{2\Omega}\textrm{Tr}\ln\bigg\{\Big[e^{\frac{\Box}{\Lambda^2_A}}\Box+q^2\hat{\phi}^2\nonumber\\
&\times&\Big(1-\frac{m^2}{\Lambda^2_\phi}\Big)\Big]\left(P_T\right)_{\mu\nu}
 +\bigg[\xi^{-1}e^{\frac{\Box}{\Lambda^2_A}}\Box+q^2\hat{\phi}^2f(\Box)\bigg]\left(P_L\right)_{\mu\nu}-\Big[e^{\frac{\Box}{\Lambda^2_\phi}}(\Box-m^2)+\frac{\lambda}{6}\hat{\phi}^2\Big]^{-1}\nonumber\\
&\times&q^2\hat{\phi}^2f^2(\Box)\partial_\mu\partial_\nu\bigg\} \ .
\eea
Since we are interested in the effective potential up to a $\hat{\phi}$-independent constant, we can factor out the operators $-e^{\frac{\Box}{\Lambda^2_\phi}}\Box\delta_{ab}$ and $e^{\frac{\Box}{\Lambda^2_A}}\Box[(P_T)_{\mu\nu}+\xi^{-1}(P_L)_{\mu\nu}]$ from the first and second traces, respectively, and then drop them. Thus, Eq. (\ref{EP}) can be rewritten as
\bea
\label{EP2}
V^{(1)}(\hat{\phi})&=&-\frac{i}{2\Omega}\textrm{Tr}\ln\bigg\{\delta_{ab}-\frac{e^{-\frac{\Box}{\Lambda^2_\phi}}}{\Box}M_{ab}\bigg\}-\frac{i}{2\Omega}\textrm{Tr}\ln\bigg\{{\delta_\mu}^\nu+q^2\hat{\phi}^2\Big(1-\frac{m^2}{\Lambda^2_\phi}\Big)\frac{e^{-\frac{\Box}{\Lambda^2_A}}}{\Box}{\left(P_T\right)_\mu}^\nu\bigg\}\nonumber\\
&-&\frac{i}{2\Omega}\textrm{Tr}\ln\bigg\{{\delta_\mu}^\nu+\xi q^2\hat{\phi}^2\bigg[1-\Big[e^{\frac{\Box}{\Lambda^2_\phi}}(\Box-m^2)+\frac{\lambda}{6}\hat{\phi}^2\Big]^{-1}\Box f(\Box)\bigg]f(\Box)\frac{e^{-\frac{\Box}{\Lambda^2_A}}}{\Box}{\left(P_L\right)_\mu}^\nu\bigg\} , \ \
\eea
where $M_{ab}\equiv\big(e^{\frac{\Box}{\Lambda^2_\phi}}m^2-\frac{\lambda}{6}\hat{\phi}^2\big)\delta_{ab}-\frac{\lambda}{3}\hat{\phi}_a\hat{\phi}_b$.

If $A$ is a diagonalizable matrix, then $\textrm{tr}f(A)=\sum_if(\lambda_i)$, where $\lambda_i$ are the eigenvalues of $A$ \cite{Higham}. Thus, in order to calculate the matrix traces in Eq. (\ref{EP2}), we need to find the eigenvalues of $M_{ab}$, ${\left(P_T\right)_\mu}^\nu$, and ${\left(P_L\right)_\mu}^\nu$. In $d$ dimensions, they are respectively
\bea
\label{Ev}
\lambda_M=\Big(e^{\frac{\Box}{\Lambda^2_\phi}}m^2-\frac{\lambda}{2}\hat{\phi}^2 \ , \ e^{\frac{\Box}{\Lambda^2_\phi}}m^2-\frac{\lambda}{6}\hat{\phi}^2\Big) \ ; \ \lambda_T=(1,1,\ldots,1,0) \ ; \ \lambda_L=(0,0,\ldots,0,1) \ .
\eea
Therefore, after some algebraic work, it can be shown from (\ref{EP2}) and (\ref{Ev}) that
\bea
\label{integrals}
V^{(1)}(\hat{\phi})&=&-\frac{i\mu^\varepsilon}{2}\int_0^\infty \frac{d^{4-\varepsilon}k}{(2\pi)^{4-\varepsilon}}\bigg\{(3-\varepsilon)\ln\bigg[1-\frac{m_A^2}{k^2}e^{\frac{k^2}{\Lambda^2_A}}\bigg]+\ln\bigg[1-\frac{m_B^2(k^2)}{k^2}e^{\frac{k^2}{\Lambda^2_\phi}}\bigg]\nonumber\\
&+&\ln\bigg[1-\frac{m_{C+}^2(k^2)}{k^2}e^{\frac{k^2}{\Lambda^2_\phi}}\bigg]+\ln\bigg[1-\frac{m_{C-}^2(k^2)}{k^2}e^{\frac{k^2}{\Lambda^2_\phi}}\bigg]\bigg\} \ , \ \
\eea 
where $\varepsilon=4-d\rightarrow0$, $\mu$ is the usual arbitrary mass scale introduced in dimensional regularization to keep the canonical dimension of $V^{(1)}(\hat{\phi})$ equals to 4, and the $m_i^2$'s are defined as follows:
\bea
m_A^2&=&q^2\hat{\phi}^2\Big(1-\frac{m^2}{\Lambda^2_\phi}\Big) \ ; \ m_B^2(k^2) \ = \ \frac{\lambda}{2}\hat{\phi}^2-m^2e^{-\frac{k^2}{\Lambda^2_\phi}} \ ; \ m_{C\pm}^2(k^2) \ = \ \frac{1}{2}\Bigg\{\Big(\frac{\lambda}{6}\hat{\phi}^2-m^2e^{-\frac{k^2}{\Lambda^2_\phi}}\Big)\nonumber \\
&\pm&\sqrt{\Big(\frac{\lambda}{6}\hat{\phi}^2-m^2e^{-\frac{k^2}{\Lambda^2_\phi}}\Big)^2-4\xi q^2\hat{\phi}^2\Big(\frac{\lambda}{6}\hat{\phi}^2-m^2\Big)e^{\big(\frac{1}{\Lambda_A^2}-\frac{1}{\Lambda^2_\phi}\big)k^2}\bigg[e^{-\frac{k^2}{\Lambda^2_\phi}}+\frac{m^2}{k^2}\Big(e^{-\frac{k^2}{\Lambda^2_\phi}}-1\Big)\bigg]}\Bigg\} . \  \  \ 
\eea
Unfortunately, the integrals in (\ref{integrals}) cannot be evaluated analytically in the most general case. Thus, to simplify the integrals, we make the assumption that $\xi=0$ (Landau gauge). Moreover, since we are interested in the Coleman-Weinberg mechanism of symmetry breaking, we assume from now on that $m^2=0$. Finally, for the sake of simplicity, we also assume that $\Lambda_A=\Lambda_\phi\equiv\Lambda$. Therefore, it follows from (\ref{integrals}) that
\bea
\label{Euclideanintegrals}
V^{(1)}(\hat{\phi})&=&\frac{(4\pi\mu^2)^\frac{\varepsilon}{2}}{16\pi^2\Gamma(2-\frac{\varepsilon}{2})}\int_0^\infty dk_Ek_E^{3-\varepsilon}\bigg\{(3-\varepsilon)\ln\bigg[1+\frac{\bar{m}_A^2}{k_E^2}e^{-\frac{k_E^2}{\Lambda^2}}\bigg]+\ln\bigg[1+\frac{\bar{m}_B^2}{k_E^2}e^{-\frac{k_E^2}{\Lambda^2}}\bigg]\nonumber\\
&+&\ln\bigg[1+\frac{\bar{m}_C^2}{k_E^2}e^{-\frac{k_E^2}{\Lambda^2}}\bigg]\bigg\} \ ,
\eea 
where we have performed a Wick rotation to Euclidean space, and the $\bar{m}_i^2$'s are defined as
\bea
\label{effmasses}
\bar{m}_A^2=q^2\hat{\phi}^2 \ ; \ \bar{m}_B^2=\frac{\lambda}{2}\hat{\phi}^2 \ ; \ \bar{m}_{C}^2=\frac{\lambda}{6}\hat{\phi}^2 \ .
\eea
Despite the huge simplification, the integrals in (\ref{Euclideanintegrals}) still cannot be evaluated exactly. For this reason, we will assume that the quantities $\bar{m}_i^2$ and $\Lambda^2$ satisfy $\Lambda^2\gg\bar{m}_i^2$. This assumption will allow us to obtain small nonlocal corrections to the Coleman-Weinberg potential. In our previous work \cite{GNPP}, we have shown that Feynman integrals similar to the ones in (\ref{Euclideanintegrals}) can be evaluated with the aid of the strategy of expansion by regions \cite{BS}. Therefore, by using this approximation method in (\ref{Euclideanintegrals}), we are able to show that
\bea
\label{unreffpot}
V^{(1)}(\hat{\phi})&=&\frac{1}{32\pi^2}\sum_{i=A,B,C}\bigg\{n_i\bar{m}_i^2\Lambda^2+\frac{1}{4}\bigg[n_i\bar{m}_i^4+2n_i\bar{m}_i^4\ln\left(\frac{2\bar{m}_i^2}{e^{1-\gamma}\Lambda^2}\right)\bigg]+\frac{n_i\bar{m}_i^6}{\Lambda^2}\ln\left(\frac{3\bar{m}_i^2}{e^{1-\gamma}\Lambda^2}\right)\nonumber\\
&+&\frac{1}{2\Lambda^4}\bigg[-n_i\bar{m}_i^8+4n_i\bar{m}_i^8\ln\left(\frac{4\bar{m}_i^2}{e^{1-\gamma}\Lambda^2}\right)\bigg]\bigg\}+\mathcal{O}\left(\Lambda^{-6}\right) \ ,
\eea
where $n_A=3$ and $n_B=n_{C}=1$.

We notice that the one-loop correction for the effective potential is finite. This ultraviolet finiteness of $V^{(1)}(\hat{\phi})$ is due to the exponential suppression of the integrals in (\ref{Euclideanintegrals}). It is important to point out that the nonlocal effects do not decouple from (\ref{unreffpot}) in the limit $\Lambda\rightarrow\infty$. However, since the local scalar QED is renormalizable, all terms that grow as $\Lambda$ grows can be absorbed into finite renormalizations of its coupling constants \cite{Burgess}. Thus, our calculation of the one-loop effective potential is an example of the applicability of the decoupling theorem \cite{AC}.

The renormalized effective potential to one-loop order is given by \cite{Schwartz}
\bea
\label{reneffpot}
V_{eff}(\hat{\phi})=\delta_\Lambda-\frac{\delta_m}{2}\hat{\phi}^2+\frac{\lambda_R+\delta_\lambda}{4!}\hat{\phi}^4+V^{(1)}(\hat{\phi}) \ ,
\eea
where $\delta_\Lambda$, $\delta_m$, and $\delta_\lambda$ are counterterms which will be used to eliminate the unwanted terms of (\ref{unreffpot}). In particular, $\delta_\Lambda$ is introduced to cancel out all additive constants dropped during the calculation.

The counterterms can be fixed by imposing the standard renormalization conditions
\bea
\label{rencond}
V_{eff}(0)=0 \ ; \ V_{eff}^{\prime\prime}(0)=0 \ ; \ V_{eff}^{\prime\prime\prime\prime}(\langle\phi\rangle)=\lambda_R \ ,
\eea
where $\langle\phi\rangle$ is the minimum of the local effective potential. Therefore, it follows from (\ref{unreffpot}-\ref{rencond}) that 
\bea
\label{finaleffpot}
V_{eff}(\hat{\phi})&=&\frac{\hat{\phi}^4}{4!}\bigg\{\lambda_R+\frac{1}{8\pi^2}\Big(9q_R^4+\frac{5}{6}\lambda_R^2\Big)\bigg[\ln\bigg(\frac{\hat{\phi}^2}{\langle\phi\rangle^2}\bigg)-\frac{25}{6}\bigg]\bigg\}-\frac{\hat{\phi}^4}{4!\Lambda^2}\sum_{i=A,B,C}\bigg\{\frac{9n_ic_i^3}{8\pi^2}\nonumber\\
&\times&\bigg[19\langle\phi\rangle^2+10\langle\phi\rangle^2\ln\bigg(\frac{3c_i\langle\phi\rangle^2}{e^{1-\gamma}\Lambda^2}\bigg)-\frac{2}{3}\hat{\phi}^2\ln\bigg(\frac{3c_i\hat{\phi}^2}{e^{1-\gamma}\Lambda^2}\bigg)\bigg]\bigg\}-\frac{\hat{\phi}^4}{4!\Lambda^4}\sum_{i=A,B,C}\bigg\{\frac{n_ic_i^4}{4\pi^2}\nonumber\\
&\times&\bigg[428\langle\phi\rangle^4+420\langle\phi\rangle^4\ln\bigg(\frac{4c_i\langle\phi\rangle^2}{e^{1-\gamma}\Lambda^2}\bigg)+\frac{3}{2}\hat{\phi}^4+6\hat{\phi}^4\ln\bigg(\frac{4c_i\hat{\phi}^2}{e^{1-\gamma}\Lambda^2}\bigg)\bigg]\bigg\}+\mathcal{O}\left(\Lambda^{-6}\right) \ ,
\eea
where $c_A=q_R^2$, $c_B=\frac{\lambda_R}{2}$, and $c_{C}=\frac{\lambda_R}{6}$.

Note that, after the renormalization, the remaining nonlocal corrections behave as $\Lambda^{-2n}$ and $\Lambda^{-2n}\ln\left(\Lambda^{-2}\right)$, so that they decouple in the limit $\Lambda\rightarrow\infty$. In the vicinity of $\langle\phi\rangle$, the nonlocal effects are given by small corrections $(\langle\phi\rangle^2\Lambda^{-2})^{n}$ and $(\langle\phi\rangle^2\Lambda^{-2})^{n}\ln\left(\langle\phi\rangle^2\Lambda^{-2}\right)$ to the local effective potential, so that these effects cannot qualitatively change the physical character of the local effective potential in the vicinity of $\langle\phi\rangle$. Thus, if we assume that $\lambda_R\sim q_R^4\ll1$, then the effective potential (\ref{finaleffpot}) will have a minimum at $\hat{\phi}_{min}\approx\langle\phi\rangle\neq0$, so that the symmetry will be spontaneously broken \cite{CW}. At the same time, the assumption $\lambda_R\sim q_R^4\ll1$ implies that we have to neglect the term proportional to $\lambda_R^2$ and all nonlocal corrections in the effective potential. Indeed, the leading nonlocal correction to (\ref{finaleffpot}) is of the order $q_R^6$, which is negligible compared to $q_R^4$. Moreover, since there are terms proportional to $q_R^6$ at two-loops and we calculated only the one-loop contributions to (\ref{finaleffpot}), then the consistency requires that we drop the nonlocal corrections in (\ref{finaleffpot}) \cite{Kang}. Therefore, if we set $\lambda_R=\frac{33}{8\pi^2}q_R^4$, then the above effective potential reduces to
\bea
\label{CWpotential}
V_{CW}(\hat{\phi})=\frac{3q_R^4}{64\pi^2}\hat{\phi}^4\bigg[\ln\bigg(\frac{\hat{\phi}^2}{\langle\phi\rangle^2}\bigg)-\frac{1}{2}\bigg]+\mathcal{O}\left(q_R^{6}\right) \ ,
\eea
which is nothing more than the usual Coleman-Weinberg potential.

At this point of the calculation, we can repeat the same analysis we did in the previous section to show that the modified dispersion relations for the fields $A_\mu(x),\sigma(x)$, and $\pi(x)$ are
\bea
\label{dr2}
e^{-\frac{p^2}{\Lambda^2}}p^2=q_R^2\langle\phi\rangle^2 \ ; \ e^{-\frac{p^2}{\Lambda^2}}p^2=V_{CW}^{\prime\prime}(\langle\phi\rangle) \ ; \ e^{-\frac{p^2}{\Lambda^2}}p^2=0 \ .
\eea
Finally, the physical solutions of (\ref{dr2}) are given by (\ref{sol}) with the following masses:
\bea
\label{ma2}
m^2_A&=&-\Lambda^2 W_0\left[-\frac{q_R^2\langle\phi\rangle^2}{\Lambda^2}\right]\approx q_R^2\langle\phi\rangle^2\left(1+\frac{q_R^2\langle\phi\rangle^2}{\Lambda^2} \right)+\mathcal{O}\left(q_R^{6}\right) \ ;\\
\label{ms2}
m^2_\sigma&=&-\Lambda^2 W_0\left[-\frac{3q_R^4\langle\phi\rangle^2}{8\pi^2\Lambda^2}\right]\approx\frac{3q_R^4\langle\phi\rangle^2}{8\pi^2}+\mathcal{O}\left(q_R^{6}\right) \ ;\\
\label{mp2}
m^2_\pi&=&0 \ ,
\eea
where we have kept terms up to the order $q_R^4$.

Note that the gauge field received a nonlocal correction to its mass, where such correction is proportional to $q_R^4\langle\phi\rangle^4$. On the other hand, the mass of the Higgs field did not receive any nonlocal correction due to the fact that the nonlocal corrections to the one-loop effective potential (\ref{finaleffpot}) are negligible. Finally, in contrast to the NLAHM, we find that Nambu-Goldstone field is massless (\ref{mp2}), so that our model (\ref{NLSQED}) is consistent with the Goldstone's theorem in the case where $m^2=0$. It is worth to point out that even if we had not dropped the nonlocal corrections in (\ref{finaleffpot}), we would still obtain the result (\ref{mp2}) because the effective potential (\ref{finaleffpot}) is a function only of the variable $\hat{\sigma}^2$.

\section{Summary}
\label{Conc}

We studied the problem of validity of the Goldstone theorem in nonlocal scalar QED. It has been argued that apparently in nonlocal theories this theorem is violated, however, up to now it was not clear what situation explicitly occurs in theories with $f(\Box)$ nonlocality. To carry out this checking, we considered the spontaneous breaking of the gauge symmetry on the tree and one-loop levels for the nonlocal QED, with the nonlocality is introduced both in gauge and scalar sectors.  We explicitly demonstrated that one of scalar fields turns out to be massless in the one-loop approximation while the non-zero masses depending on constant background field arise for other fields, in other words, the Goldstone theorem is fulfilled at the one-loop level but not at the tree level, except of the special case where the scalar field is massless, hence, the problem of validity of the Goldstone theorem appears to be highly controversial. In principle, the massless case requires more careful studies. Effectively, here we give the first constructive example of checking the Goldstone theorem for nonlocal theories. Also, it deserves to mention that, within our study we, for the first time, explicitly calculated the Coleman-Weinberg effective potential for the nonlocal case, while in the previous paper \cite{Bri} by some of us, it has been evaluated for the nonlocal theory involving only the scalar field.

As a possible continuations of this study, it would be natural to suggest to consider this calculation at the finite temperature case with a subsequent study of possibility of phase transitions. We suggest to do it in our next paper. 

{\bf Acknowledgements.} This work was partially supported by Conselho Nacional de Desenvolvimento Cient\'{i}­fico e Tecnol\'{o}gico (CNPq). The work by A. Yu. P. has been partially supported by the CNPq project No. 303783/2015-0.

\end{document}